\documentclass[twocolumn,showpacs,preprintnumbers,amsmath,amssymb,superscriptaddress]{revtex4}

\usepackage{graphicx}
\usepackage{rotating,epstopdf}
\usepackage{dcolumn}
\usepackage{bm}
\usepackage{color}

\begin{document}

\title{New Measurement of the Relative Scintillation Efficiency of Xenon Nuclear
Recoils Below 10 keV}

\author{E.~Aprile}
\affiliation{Department of Physics, Columbia University, New York, NY 10027,
USA}

\author{L.~Baudis}
\affiliation{Physik-Institut, Universit\"{a}t Z\"{u}rich, Z\"{u}rich, 8057,
Switzerland}

\author{B.~Choi}
\affiliation{Department of Physics, Columbia University, New York, NY 10027,
USA}

\author{K.~L.~Giboni}
\affiliation{Department of Physics, Columbia University, New York, NY 10027,
USA}

\author{K.~Lim}
\affiliation{Department of Physics, Columbia University, New York, NY 10027,
USA}

\author{A.~Manalaysay}
\email[Corresponding author, electronic address: ]{aaronm@physik.uzh.ch}
\affiliation{Physik-Institut, Universit\"{a}t Z\"{u}rich, Z\"{u}rich, 8057,
Switzerland}
\affiliation{Department of Physics, University of Florida, Gainesville, FL
32611, USA}

\author{M.~E.~Monzani}
\affiliation{Department of Physics, Columbia University, New York, NY 10027,
USA}

\author{G.~Plante}
\affiliation{Department of Physics, Columbia University, New York, NY 10027,
USA}

\author{R.~Santorelli}
\affiliation{Department of Physics, Columbia University, New York, NY 10027,
USA}

\author{M.~Yamashita}
\affiliation{Department of Physics, Columbia University, New York, NY 10027,
USA}

\received{29 September, 2008}

\begin{abstract}
Liquid xenon is an important detection medium in direct dark matter experiments,
which search for low-energy nuclear recoils produced by the elastic scattering
of WIMPs with quarks.  The two existing measurements of the relative
scintillation efficiency of nuclear recoils below 20 keV lead to inconsistent
extrapolations at lower energies.  This results in a different energy scale and
thus sensitivity reach of liquid xenon dark matter detectors.  We report a new
measurement of the relative scintillation efficiency below 10 keV performed with
a liquid xenon scintillation detector, optimized for maximum light collection.
Greater than 95\% of the interior surface of this detector was instrumented with
photomultiplier tubes, giving a scintillation yield of 19.6 photoelectrons/keV
electron equivalent for 122 keV gamma rays.  We find that the relative
scintillation efficiency for nuclear recoils of 5 keV is 0.14, staying constant
around this value up to 10 keV.  For higher energy recoils we measure a value
of 0.21, consistent with previously reported data.  In light of this new
measurement, the XENON10 experiment's upper limits on spin-independent 
WIMP-nucleon cross section, which were calculated assuming a constant 0.19 
relative scintillation efficiency, change from $8.8\times10^{-44}$ cm$^2$ to
$9.9\times10^{-44}$ cm$^2$ for WIMPs of mass 100 GeV/c$^2$, and from
$4.5\times10^{-44}$ cm$^2$ to $5.6\times10^{-44}$ cm$^2$ for WIMPs of mass 30
GeV/c$^2$.
\end{abstract}


\maketitle
\newcommand{\Leff}{$\mathcal{L}_{eff}$}

\section{Introduction}
\label{sec:intro}

Numerous observations point to the existence of a non-luminous, non-baryonic
component of our universe known as dark matter
\cite{Begeman:91,Spergel:07,Clowe:06}.  This matter could be in the form of a
new type of particle \cite{Jungman:96,Bertone:05} whose existence can be
naturally explained as a thermal relic of the Big Bang.  Such a particle, which
would likely be massive and electrically neutral, is commonly referred to as a
Weakly Interacting Massive Particle, or WIMP.  There are currently a number of
efforts, worldwide, to directly detect WIMPs in terrestrial particle detectors
\cite{Gaitskell:04,Baudis:06,ZepII,XENON10,XENON10_SD}.  These are sensitive to
nuclear recoils below 100 keV \cite{PSS,LS} and take advantage of the fact that
WIMPs, interacting primarily with atomic nuclei, would be distinguishable from
the predominantly electromagnetic backgrounds present in such experiments.

While there are varying strategies in this direct detection search, liquid xenon
(LXe) has recently come to the forefront of the field as a powerful detection
medium \cite{ZepII,XENON10,XENON10_SD}.  When particles interact in LXe, they
produce prompt scintillation photons and ionization \cite{Doke:90} which can be
measured simultaneously to infer the energy and the type of interaction. The
scintillation emission spectrum from LXe is singly-peaked at 178 nm with a width
of 13 nm \cite{Jortner:65}.  The \emph{scintillation yield}, defined as the
number of photons produced per unit energy, depends on the identity of the
particle depositing the energy \cite{Lindhard,Hitachi:BiEx}.  Since WIMPs
primarily interact with atomic nuclei, in LXe searches  the quantity of interest
is the scintillation yield of recoiling xenon nuclei.  The recent XENON10
measurement \cite{XENON10,XENON10_SD} has pushed the energy threshold of LXe
detectors down to regions where the nuclear recoil scintillation yield is poorly
understood.  This lack of knowledge is the source of XENON10's largest
systematic uncertainty.

Measurement of the absolute scintillation yield of any particle species is quite
difficult, so instead relative yields are often reported.  Monoenergetic gamma
rays of 122 keV from $^{57}$Co are commonly used to calibrate the electronic
recoil energy scale.  The scintillation yield of nuclear recoils \emph{relative}
to that of 122 keV gamma rays is known as the relative scintillation efficiency,
or \Leff, and a new measurement of this energy-dependent value is reported
herein.  While the absolute scintillation yield of electronic recoils is not
linear in energy \cite{Yamashita:04}, the 122 keV gamma rays provide an anchor
point upon which one can base scintillation yields of all other energies and
species.

\section{Experimental Setup}
\label{sec:exp_apparatus}
\subsection{Neutron Beam}
\label{sec:exp_apparatus:facility}

\begin{figure}[htb]
	\begin{center}
		\includegraphics[width=0.45\textwidth]{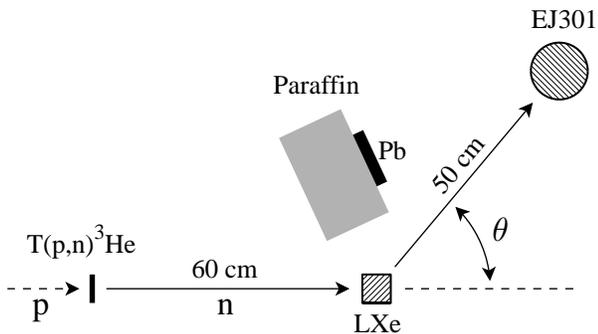} 
	\end{center}
	\caption{Schematic diagram of the experimental setup.  Incoming 1 MeV 
	neutrons scatter in the LXe detector and are tagged by the EJ301 neutron
	scintillator, for different scattering angles of 48$^{\circ}$, 62$^{\circ}$,
	70.5$^{\circ}$, and 109.5$^{\circ}$.  The paraffin and lead are used to
	shield the EJ301 neutron scintillator from direct neutrons and gamma rays.}
	\label{fig:apparatus}
\end{figure}

The determination of \Leff~requires the production of nuclear recoils whose
energies are known independent of their response in the LXe.  Nearly
monoenergetic neutrons, produced in a T(p,n)$^3$He reaction, are incident upon a
LXe detector. Some of the neutrons which scatter under an angle $\theta$ are
detected by an EJ301 organic liquid scintillator detector \cite{Eljen} (see
Fig.~\ref{fig:apparatus}), capable of distinguishing electronic (gamma rays)
from nuclear (neutron) recoils via Pulse Shape Discrimination (PSD)
\cite{Knoll,Marrone}.  In this way, the energy of the recoiling Xe nucleus is
known kinematically, and is given by the relation

\begin{equation}
E_r \approx 2E_n \frac{m_nM_{N}}{(m_n+M_{N})^2}(1 - \mathrm{cos}\,\theta),
\end{equation}

\noindent where $E_r$ is the recoil energy, $E_n$ is the energy of the incoming
neutron, $m_n$ and $M_{N}$ are the masses of the neutron and nucleus,
respectively, and $\theta$ is the scattering angle (the approximation is valid
when $M_{N} \gg m_n$ and $E_{n} \ll m_{n}c^{2}$).

The measurements were conducted in the neutron beam of the Radiological Research
Accelerator Facility at the Columbia Nevis Laboratory, also described in
a previous study of \Leff~\cite{Aprile:05}.  In the present work, 1.9 MeV
protons were incident upon a tritium target, yielding 1 MeV neutrons.  The
T(p,n)$^3$He reaction produces neutrons over 4$\pi$ sr, however, the luminosity
is peaked in the forward direction and the energy variation due to the angular
spread of the 1'' LXe detector cell, 60 cm distant from the tritium target, was
less than 0.09\% \cite{PN_tables:73}.  The incident proton energy, $E_p$, was
known to within 0.1\%. These two systematic uncertainties, coming from the
angular dependence of $E_n$ and the uncertainty in $E_p$, are considered
negligible and are not included in the calculations of section
\ref{sec:analysis:cuts_BG}.  The dominant spread in the incident neutron energy
comes from the thickness of the tritium target, and was estimated to give a
1-$\sigma$ spread of $\pm$7.8\% \cite{Marino}.  An additional component to the
spread in $E_r$ comes from the finite size of both detectors, which leads to an
uncertainty in the true scattering angle, $\theta$.  These uncertainties were
determined by Monte Carlo (MC) simulations, and are included in the
uncertainties in Table \ref{tab:results}.

Also seen in Fig.~\ref{fig:apparatus}, a 30 cm-thick paraffin block was placed
along the line of sight between the tritium target and the EJ301 scintillator,
in order to block neutrons from directly interacting in the EJ301.  In addition
to the paraffin block, 5 cm of Pb shielded the EJ301 from gammas produced in the
target.  The data presented here were accumulated in several data sets with the
following neutron scattering angles: 48$^{\circ}$, 62$^{\circ}$, 70.5$^{\circ}$,
and 109.5$^{\circ}$.

\subsection{Detector Design}
\label{sec:exp_apparatus:detector}

\begin{figure}[htp!]
	\includegraphics[width=0.48\textwidth]{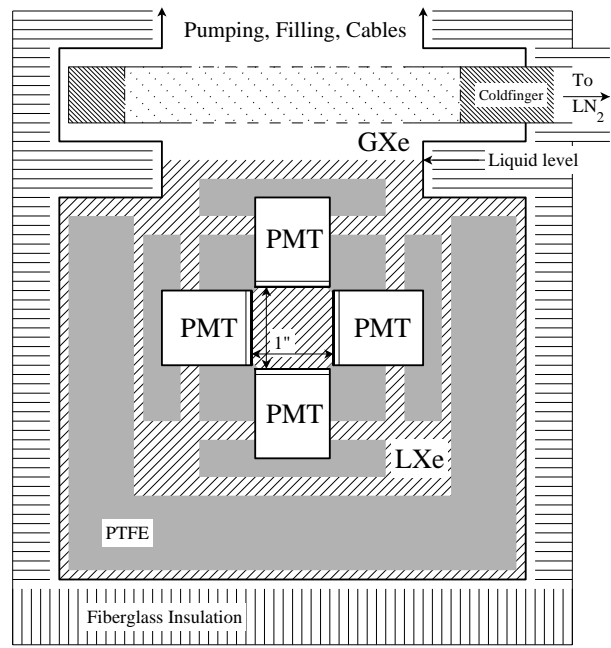} 
	\caption{Schematic diagram of the LXe detector used for the
	\Leff~measurement.  Visible are four of the six PMTs used to view the 1
	in$^3$ active LXe volume.}
	\label{fig:detector}
\end{figure}

The LXe detector was designed to allow a zero-field measurement of the
scintillation light with very high efficiency, covering $>$95\% of the interior
surface with photon detectors.  The design also minimizes the amount of LXe
outside the active volume, to reduce the background from interactions in this
passive scintillation layer.  A schematic of the detector design is seen in
Fig.~\ref{fig:detector}.  The LXe volume is viewed by six 1''-square Hamamatsu
metal channel R8520-06-Al photomultiplier tubes (PMTs), sensitive to the 178 nm
Xe scintillation.  Four of these PMTs use a new bialkali photocathode, with a
quantum efficiency around 40\% at room temperature \cite{Hamamatsu}.  The PMTs,
held together with a polytetrafluoroethylene (PTFE) frame, form a cube such that
each PMT window covers a face of the cube.  Both the photocathode and metal body
of the PMTs are held at ground potential, with positive high voltage applied to
the anodes.  This configuration guarantees that no residual electric fields
exist in the LXe, whose scintillation yield strongly depends on the applied
field \cite{Aprile:06} (by definition, \Leff~is the relative light yield at zero
field). The PMT assembly is mounted in a stainless steel vessel, surrounded by
fiberglass for thermal insulation.

Xe gas, purified by the same system used in \cite{Aprile:05}, is liquefied in
the vessel, cooled by a copper ring cold finger thermally coupled to a liquid
nitrogen bath.  The LXe level is kept above the top PMT and the temperature was
held constant at 180K (the same T and P as in XENON10
\cite{XENON10,XENON10_SD}), with fluctuations less than 0.03\%.  A total of 2.5
kg of Xe was used.

The EJ301 scintillator is contained in an aluminum cylinder 3" in diameter and
3" tall, viewed by a single Photonis XP4312B PMT and read out with the same
electronics as the PMTs in the LXe detector.

\subsection{Data Acquisition and Processing}
\label{sec:exp_apparatus:daq}


The PMT signals are fed into an amplifier, with two identical outputs per intput
channel.  One output is digitized by a 100 MHz flash ADC, while the other output
is fed to the triggering system.  

For the trigger, the six LXe PMT channels are combined in pairs to form three
trigger channels, each discriminated at a level of 0.3 photoelectrons (p.e.).
The logical outputs of the three discriminator channels are passed to an $N=3$
coincidence unit.  The efficiency of this trigger condition, determined by the
Monte Carlo method, is $\sim$100\% at 20 p.e., slowly rolling off to $\sim$90\%
at 10 p.e.  The EJ301 trigger is taken simply as the output of the discriminated
signal.  

For the measurement of the neutrons' Time of Flight (ToF) the LXe trigger is fed
directly to the ``start'' input of a Time-to-Amplitude Converter (TAC), while 
the EJ301 trigger provides the ``stop'' after appropriate delay.  The
output of the TAC is digitized by the same flash ADC unit.  Calibration of the 
ToF signal is discussed further in section \ref{sec:analysis:calibration}.

The shape of the signal in the EJ301 depends on the incoming particle species,
and can be used to distinguish neutrons from gamma rays since the characteristic
scintillation decay time is different for these particles.  As a result, the
tails of pulses resulting from nuclear recoils will be characteristically longer
than those from electronic recoils.  A description of the mechanisms involved in
this process can be found in Ref.~\cite{Knoll}.  In EJ301, the ``slow''
component is two orders of magnitude longer than the ``fast'' component,
reported to be 3.2 ns \cite{Eljen}.  A PSD parameter is constructed by dividing
the area under the pulse's tail by the total area of the pulse, with the tail
defined as the part of the trace starting 30 ns after the peak until the trace
reaches 5\% of the peak value.

\section{Data Analysis and Results}
\label{sec:analysis}
\subsection{Calibrations}
\label{sec:analysis:calibration}

The PMTs were calibrated \emph{in situ} with a pulsed blue LED, in order to
measure and monitor the gain.  The light from the LED produces a single
p.e.~spectrum, whose mean determines the gain of the multiplier chain.  With a
complete set of such LED calibration measurements, the signals obtained for all
acquisitions can be converted to a value in number of p.e.  The relationship
between the number of collected p.e.~and the total number of emitted photons
depends on the geometrical light collection efficiency, the quantum efficiency
of the photocathodes, and the collection efficiency between the photocathode and
the first dynode.  Although these values are not known to very high precision,
they represent completely linear processes and hence lead to a linear
relationship between the total number of scintillation photons and the measured
number of p.e.  Comparing the p.e.~yields of various sources thus gives a
measure of their relative scintillation yields.

\begin{figure}[h]
	\begin{center}
		\includegraphics[width=8.6cm]{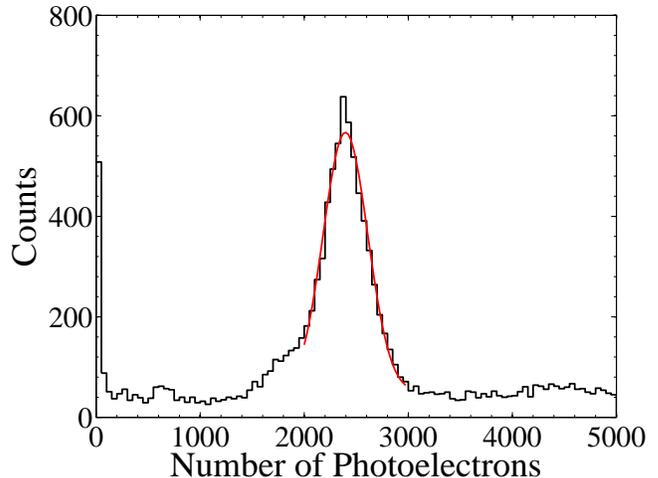} 
	\end{center}
	\caption{(color online).  The scintillation light spectrum of 122 keV gamma
	rays from $^{57}$Co, used to calibrate the electronic recoil energy scale.
	This calibration gives a scintillation yield of 19.64 p.e./keV.}
	\label{fig:Co57}
\end{figure}

As \Leff~is defined relative to the scintillation yield of 122 keV gamma rays,
data from a 100 $\mu$Ci $^{57}$Co source were taken periodically during the
experiment.  Fig.~\ref{fig:Co57} shows the spectrum from one such calibration.
The $^{57}$Co yield was measured to be 19.64 $\pm$ 0.07 (stat) $\pm$ 0.11 (sys)
p.e./keVee, where the statistical uncertainty is the combination of the
parameter uncertainties of the fits from the various calibration data, and the
systematic uncertainty is taken from the variation in this yield over the
two-day duration of the experiment.  One set of PMT gain values was applied to
all data, and thus the systematic uncertainty in the $^{57}$Co yield quoted
above accounts for both variations in yield and PMT gain.

In addition to $^{57}$Co, data were also collected from a $^{22}$Na source.
This source emits a $\beta^{+}$ which promptly loses energy in the Na and
annihilates, producing two 511 keV gamma rays emitted simultaneously in opposite
directions.  With the source placed between the LXe detector and the EJ301
detector, the two gamma rays will interact at essentially the same time in the
two detectors.  In this way, $^{22}$Na provides a baseline ToF=0 which, when
used in conjunction with a variable delay generator, is used to calibrate the
ToF measurement system.

\subsection{Event Selection, Backgrounds and Results}
\label{sec:analysis:cuts_BG}

\begin{figure}[htp]
		\includegraphics[width=8.6cm]{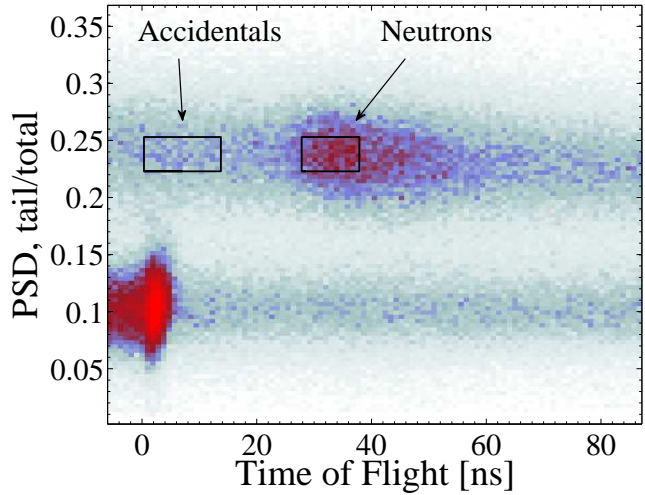} 
	\caption{(color online).  The distribution of triggered events in PSD
	vs.~ToF space from the data set at 70.5$^{\circ}$.  An ``upper'' band and
	``lower'' band are readily identifiable in the data, and correspond to
	nuclear recoils and electronic recoils, respectively.  The peak at the lower
	left near ToF=0, due to gamma rays that Compton scatter in the LXe before
	striking the EJ301, is easily vetoed by the PSD cut.  A population of
	accidental triggers (see text) having a flat ToF spectrum is visible in both
	bands and contributes background events within the neutron peak.  The LXe
	spectra of events within the left box are used as the expectations of this
	background.  The width of the right box---10 ns---is chosen to accept
	neutrons that interact in any region of the finitely-sized detectors.}
	
	\label{fig:PSD_ToF}
\end{figure}

The processing of the data acquired at each angle yields two parameters which
can be used to select events of interest: the event ToF, and the PSD parameter
from the EJ301 neutron detector.  Fig.~\ref{fig:PSD_ToF} shows the distribution
of events in PSD parameter and ToF.  Clearly visible are the nuclear recoil and
electronic recoil bands, in addition to the peaks from both gamma and neutron
scatters.  The PSD cut is chosen to accept a majority of the nuclear recoil band
while rejecting electronic recoils.  The width of the ToF cut is 10 ns, which is
the expectation based on the spread in $E_n$ and the finite size of the
detectors.  The tail of the ToF peak is due mainly to events where the neutron
scattered in one of the detector materials in addition to the LXe, before
interacting in the EJ301 scintillator.  Multiple scatters in the LXe also add to
the tail, although MC simulations indicate that their overall contribution is
less than 2\%.

\begin{figure*}[htp!]
	\begin{center}
		\includegraphics[width=17.2cm]{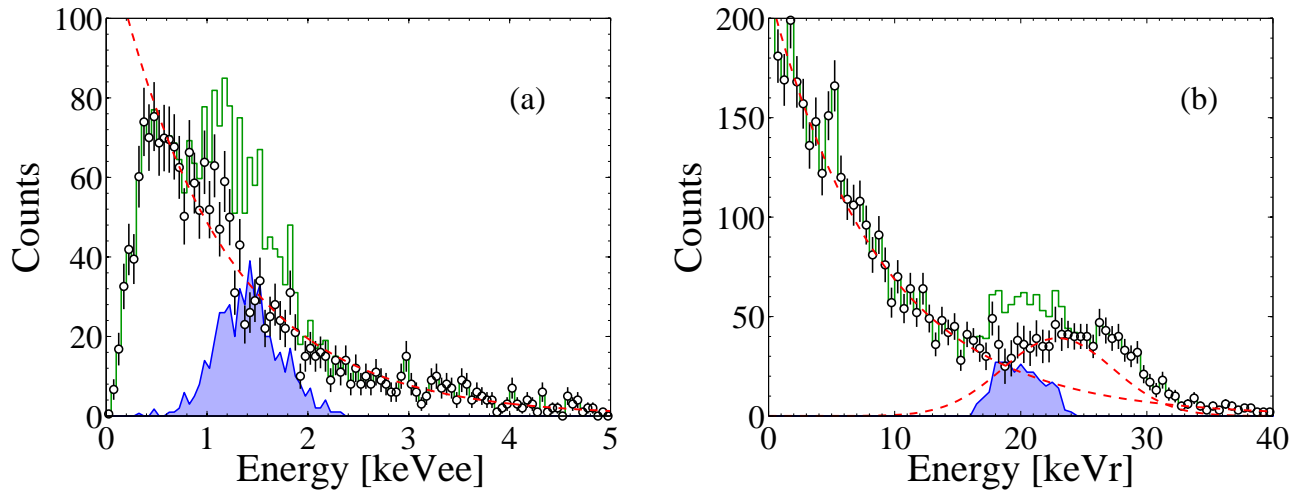}
	\end{center}
	\caption{(color online). 
	Selected results of the Monte Carlo simulations, which do not
	include the accidentals background.  (a)--The spectrum of events tagged at
	70.5$^{\circ}$, scaled with the measured value of \Leff~giving the 
	electron-equivalent energy (keVee), convoluted assuming Poisson statistics 
	for the number of p.e.~and multiplied by the simulated trigger efficiency
	curve.  
	The green histogram is the total spectrum, and the black circles indicate 
	the true materials background.  The red dashed line is an exponential fit to 
	the high-energy region of the green histogram; its agreement with the true 
	materials background confirms the validity of this technique's use in the 
	real data.  The shaded blue area shows the spectrum of true elastic single 
	scattered neutrons.  (b)--The spectrum of events tagged at 109.5$^{\circ}$.  
	The data are shown in the original, recoil equivalent energy scale (keVr) 
	without Poisson convolution.  The materials background in this region 
	departs from the exponential behavior seen at lower energies, and distorts 
	the position of the peak from true single scatters, at 20 keV.  The red 
	dashed lines are the result of an exponential+Gaussian fit.  The Gaussian 
	component, centered at $22.94\pm4.34$ keV, is used as the `true' energy of 
	the Gaussian component in the real spectrum.}
	\label{fig:MC_spects}
\end{figure*}

Two backgrounds contribute to the LXe spectrum which cannot be vetoed with the
cuts described above, and must instead be subtracted.  It is clear from
Fig.~\ref{fig:PSD_ToF} that beneath the neutron peak lies a population of events
which have a flat ToF spectrum.  These are identified as neutrons that
accidentally interacted in the EJ301 in coincidence with an unrelated event in
the LXe, and are referred to as accidentals.  As these events are uniform in ToF
space, accidentals outside of the ToF peak should have the same energy spectrum
as those within the peak.  The LXe spectrum of the events inside the box of
Fig.~\ref{fig:PSD_ToF} labeled ``accidentals'' was used as the
expectation of the accidentals background.  The region to the left of the peak
was chosen because the peak's extended tail contaminates the accidentals
spectrum to the right of the ``neutron'' peak.

The second background that cannot be vetoed comes from neutrons that scattered
in various detector materials in addition to the LXe, before interacting in the
EJ301.  Here referred to as materials background, MC simulations show that the
spectrum of these events in the LXe follows an approximately exponential
distribution in the region of the peak.  Fig.~\ref{fig:MC_spects}(a) displays
the results of the MC simulation of the data set at 70.5$^{\circ}$, indicating
the contribution from the materials background.  In order to estimate the
spectrum of these events in the real data, a decaying exponential was fit to the
high energy portion of the distributions after subtracting the accidentals 
background.

\begin{figure*}[htp!]
	\begin{center}
		\includegraphics[width=17.2cm]{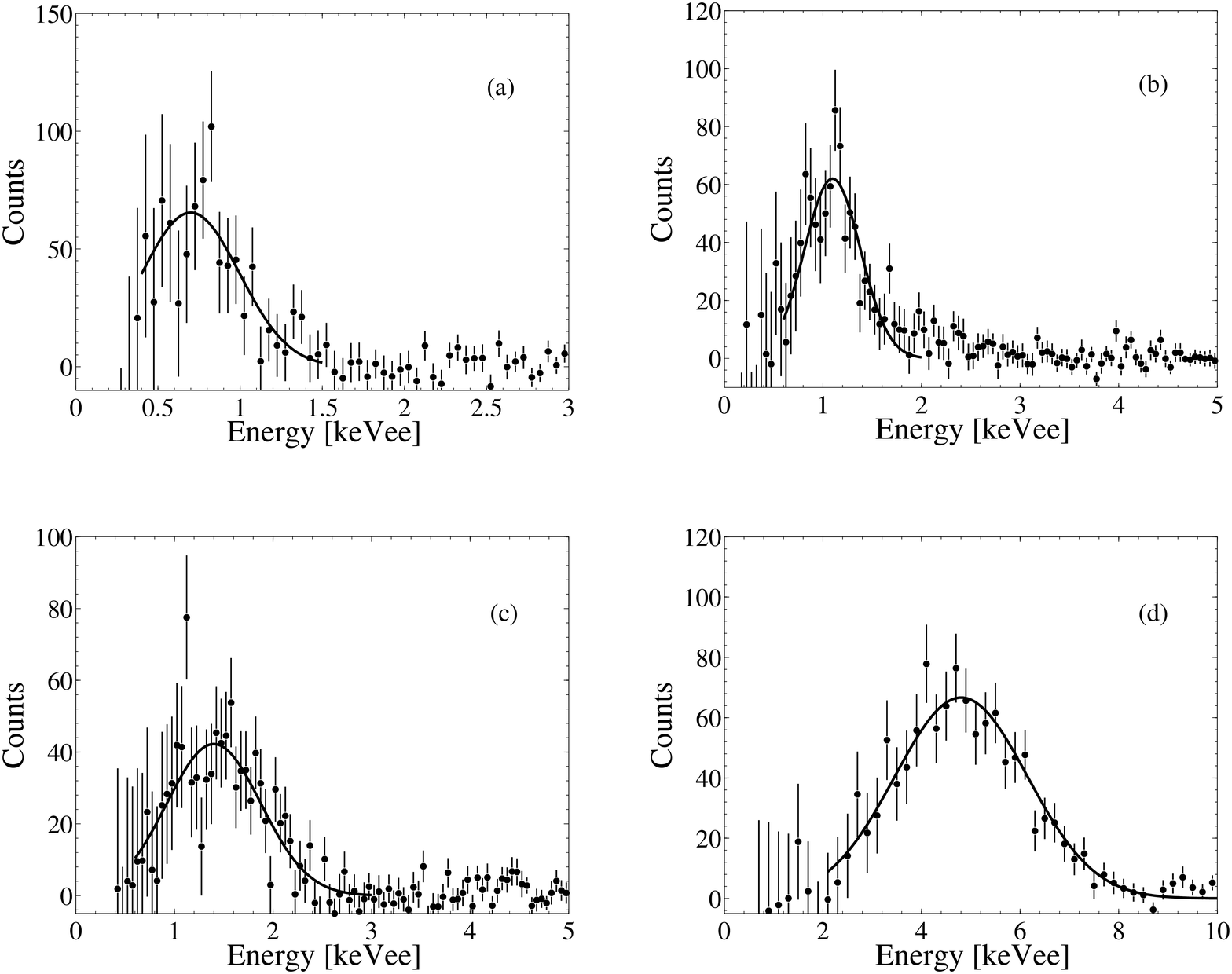}
	\end{center}
	\caption{
	Energy spectrum measured at the four angles used in this study: 
	(a)-48$^{\circ}$; (b)-62$^{\circ}$; (c)-70.5$^{\circ}$;
	(d)-109.5$^{\circ}$.  The data are shown after subtracting both 
	accidentals and materials backgrounds.  Error bars are the combined errors 
	of the original spectra, accidentals, and materials background, included in 
	the Gaussian fits indicated by the solid curves.}
	\label{fig:LXe_spectra}
\end{figure*}

After applying cuts (PSD and ToF) and subtracting backgrounds (accidentals and
materials), a spectrum results in which the peak from single-scatter neutrons
can be readily identified, seen as the solid circles in
Fig.~\ref{fig:LXe_spectra}.  The horizontal scale of these spectra is given as
``keVee'' meaning ``keV electron-equivalent'', indicating it is the energy scale
derived from the $^{57}$Co calibration.  \Leff~is found from the following
relation:

\begin{equation}
\mathcal{L}_{eff} = \frac{E_{ee}}{E_{nr}},
\end{equation}

\noindent where $E_{ee}$ is the electron-equivalent energy (based on the 122 keV
scintillation yield) and $E_{nr}$ is the true recoil energy.  Thus, when these
spectra are fit with Gaussian functions, the estimators of the mean, divided by
the true recoil energy, give the \Leff~values at these energies.

The uncertainties in the recoil energies are taken directly from the spread in
the incident neutron energy combined with the geometrical uncertainty due to the
finite size of the detectors.  These values were obtained from the MC
simulations and are listed in the second column of Table \ref{tab:results}.  The
uncertainties in \Leff~were calculated by considering the spread in $E_r$
mentioned above, statistical errors in the Gaussian fits, the variation in
$^{57}$Co light yield, the uncertainty in the background estimations, and the
effect of the trigger threshold roll-off.  This last uncertainty was calculated
by finding the peak positions before and after dividing the spectra by the
trigger efficiency discussed in section \ref{sec:exp_apparatus:daq}.  However,
only the lowest angle (48$^{\circ}$)~was affected by this trigger roll-off.  The
asymmetric error bars of the 5 keV data point is due to both the trigger roll
off and the actual parameter uncertainty in the Gaussian fit.  For all angles,
the dominant contribution to the uncertainty in \Leff~is from the spread in
$E_r$.

\begin{table}[h]
\begin{center}
\caption{\label{tab:results} The values of \Leff~obtained at the four angles
used in this study.  Error bars on the recoil energies are the spread of $E_n$
as mentioned in section \ref{sec:exp_apparatus:facility} combined with the
geometrical uncertainties.  The uncertainties in \Leff~are the combination of
all statistical and systematic errors mentioned in the text.}
\begin{ruledtabular}
	\begin{tabular}{ r  r  l }
		\multicolumn{1}{c}{$\theta$} 
		& $E_r$ (keV)
		& \multicolumn{1}{c}{\Leff} \\
		\hline 
		48$^{\circ}$ & $5\pm0.68$ & $0.141^{+0.025}_{-0.037}$ \\
		62$^{\circ}$ & $8\pm0.91$ & $0.137\pm0.016$ \\
		70.5$^{\circ}$ & $10\pm1.06$ & $0.140\pm0.016$ \\
		109.5$^{\circ}$ & $22.94\pm4.34$ & $0.205\pm0.039$ \\
	\end{tabular}
\end{ruledtabular}
\end{center}
\end{table}

\begin{figure}[htp]
		\includegraphics[width=8.6cm]{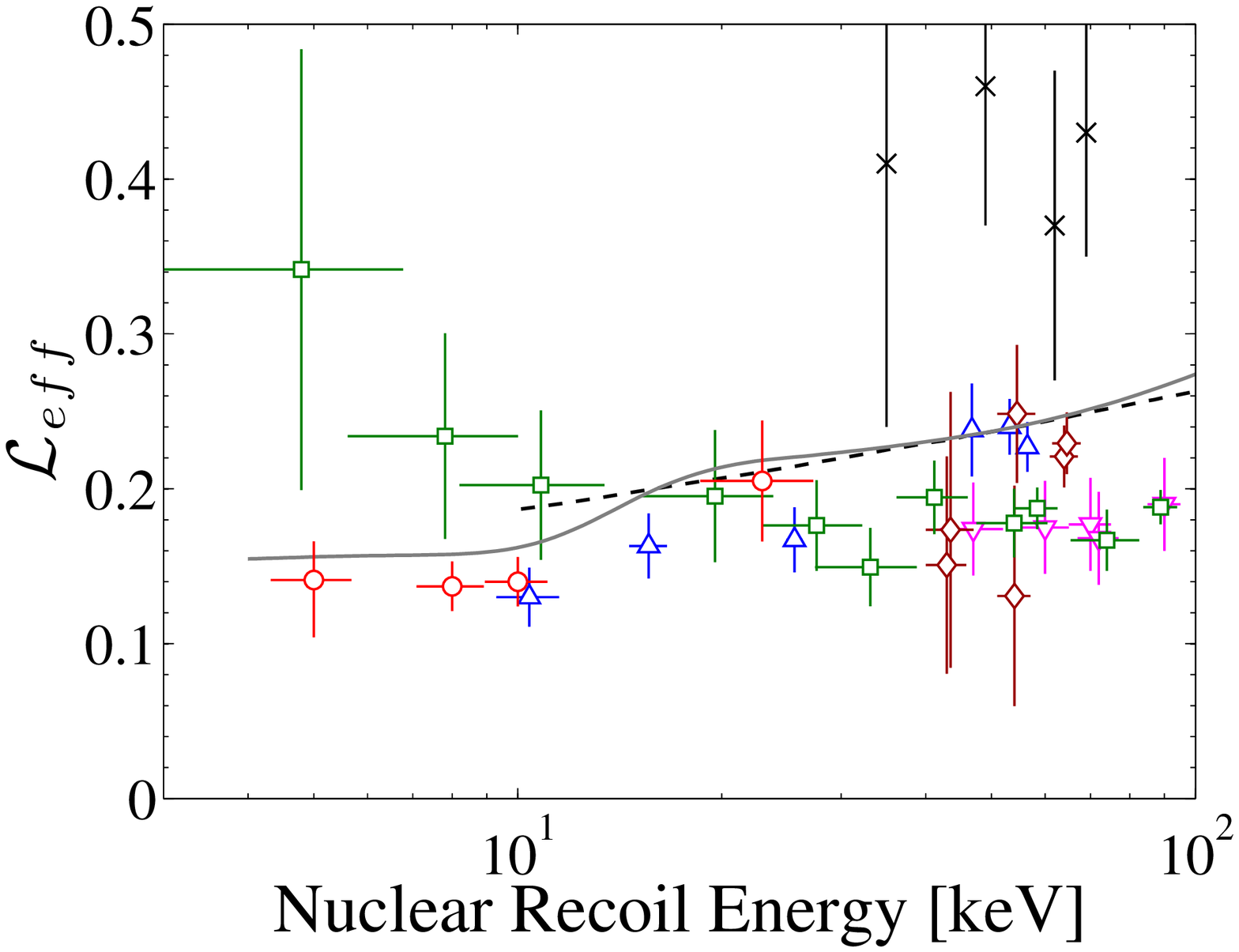} 
	\caption{(color online). 
	Measured \Leff~values as a function of Xe nuclear recoil energy.  Symbols 
	correspond to (\textcolor{red}{$\circ$})--this work;
	(\textcolor[rgb]{0,.5,0}{\bf$\square$})--Chepel et al.~\cite{Chepel:06};
	(\textcolor{blue}{$\bigtriangleup$})--Aprile et al.~\cite{Aprile:05};
	(\textcolor[rgb]{.6,0,0}{$\lozenge$})--Akimov et al.~\cite{Akimov:02};
	($\times$)--Bernabei et al.~\cite{Bernabei:01};
	(\textcolor[rgb]{1.,0,1.}{$\bigtriangledown$})--Arneodo et
	al.~\cite{Arneodo:00}.  The solid gray curve is the result from a recent
	best-fit analysis of XENON10 AmBe source data and MC \cite{Sorensen:09}.  
	Also shown is the theoretical prediction of Hitachi
	(dashed line) \cite{Hitachi:BiEx}.}
	\label{fig:Leff_plot}
\end{figure}

Though the purpose of this study was to investigate the behavior of \Leff~below
10 keV, it was necessary to collect data from higher-energy recoils in order to
establish a connection with previous studies.  For this, the EJ301 was placed at
a scattering angle of 109.5$^{\circ}$, corresponding to 20.0 keV recoils.
However, this angle is close to the minimum in the differential scattering cross
section of 1 MeV neutrons in Xe \cite{ENDF_base}, and so the signal from
``true'' single scatters is well below the background.  Additionally, the
materials background in this energy range departs from a decaying exponential.
As can be seen in the MC data of Fig.~\ref{fig:MC_spects}(b), the actual
``bump'' in the spectrum, coming primarily from neutrons which have also
scattered in the PTFE, is actually slightly higher than 20 keV.  In order to
find the true energy of the peak position, the same procedure used in examining
the real data was applied here to the MC data, giving a recoil energy of
$22.94\pm4.34$ keV.  The spread in $E_r$ was taken as the width of the Gaussian
component in the MC spectrum.

The values obtained for \Leff~are listed in Table \ref{tab:results}, and
additionally shown in Fig.~\ref{fig:Leff_plot} along with the results of
previous studies \cite{Chepel:06,Aprile:05,Akimov:02,Bernabei:01,Arneodo:00}.

\section{Discussion of Results}
\label{sec:discussion}

The data point from the measurement at 109.5$^{\circ}$~shows agreement with
other measurements whose high-energy behavior averages out to $\mathcal{L}_{eff}
\approx 0.19$.  Below 10 keV, our values are substantially lower than the
central values of Chepel et al.~\cite{Chepel:06}, with a considerable
improvement in precision.  The central value at 10 keV is consistent with the
lowest-energy data point of Aprile et al.~\cite{Aprile:05}, enforcing the
accuracy of this measurement.  Unfortunately, the theoretical models of neither
Lindhard \cite{Lindhard} nor Hitachi \cite{Hitachi:BiEx} can shed any light on
the behavior of \Leff~in this energy range.  Hitachi's model, which attempts to
take into account incomplete charge recombination and additional electronic
quenching, is based on Lindhard quenching as well as the Thomas-Fermi
approximation; for Xe nuclear recoils, both break down below 10 keV
\cite{Hitachi:07,Mangiarotti:07}.

\begin{figure}[htp]
	\begin{center}
		\includegraphics[width=8.6cm]{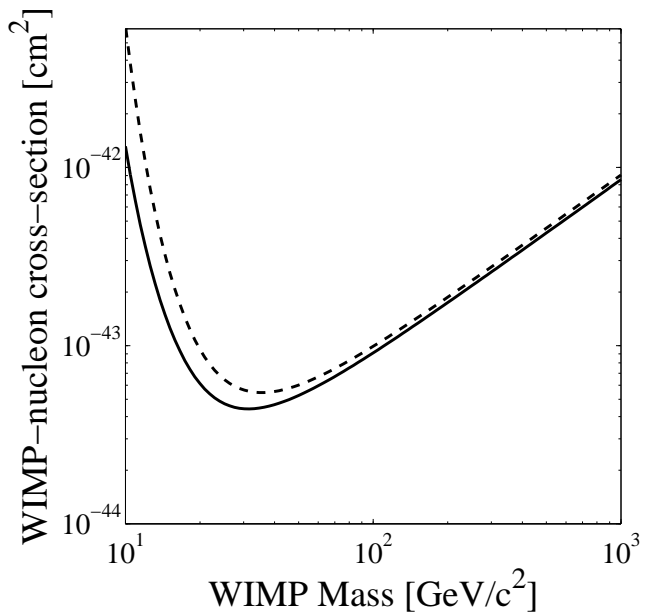} 
	\end{center}
	\caption{
	The upper limit on the WIMP-nucleon spin-independent cross section based on 
	the 58.6 live days of XENON10's WIMP search, shown with a 
	flat \Leff$=0.19$ (solid).  An \Leff~function consistent with the 
	results of this study, applied to the same XENON10 data is shown as well
	(dashed).}
	\label{fig:SI_limit}
\end{figure}

As mentioned in the introduction, the uncertainty in \Leff~at low recoil
energies presents the largest systematic uncertainty in the results of the
XENON10 dark matter experiment, where it was chosen to use a flat
$\mathcal{L}_{eff} = 0.19$ as a compromise between the seemingly opposing trends
observed by Chepel and Aprile.  Under this assumption, the WIMP-nucleon
spin-independent cross section for WIMPs of mass 100 GeV/c$^2$ was constrained
to be less than $8.8\times10^{-44}$ cm$^2$ and $4.5\times10^{-44}$ cm$^2$ at 30 
GeV/c$^2$, indicated by the solid blue curve in Fig \ref{fig:SI_limit}.
Allowing for \Leff~scenarios below 20 keV that cover the values 
allowed by both Chepel and Aprile gives upper limits that vary by $\sim$40\% at 
30 GeV/c$^2$ and $\sim$18\% at 100 GeV/c$^2$, with variations becoming less 
severe with increasing WIMP mass.  With an \Leff~model that follows the new data  
points of this study, the resulting upper limit is shown in Fig 
\ref{fig:SI_limit} as the blue dashed 
curve.  The limit is shifted up to $9.9\times10^{-44}$ cm$^2$ and 
$5.6\times10^{-44}$ cm$^2$ for WIMPs of mass 100 GeV/c$^2$ and 30 GeV/c$^2$, 
respectively.  

It has become clear from XENON10 that future dark matter searches using LXe must 
have sensitivity to nuclear recoils below 10 keV in order to be competitive.  
The improved understanding of \Leff's behavior presented in this study not only 
permits a more precise interpretation of XENON10's results, but benefits future 
dark matter searches also using LXe.  Several next generation LXe dark 
matter searches are currently in operation or under construction, such as 
XENON100 \cite{XENON100}, LUX 
\cite{LUX} and XMASS \cite{XMASS}.  These experiments will begin to probe for 
the first time those regions of parameter space most favored by many theoretical 
models, and will consequently rely quite heavily on a precise understanding of 
LXe's scintillation efficiency for low energy nuclear recoils when interpreting 
their results.  This is true in the case of a null result and especially in the 
case of a positive signal.  If and when such a signal is detected, a measurement 
of the WIMP mass, for example, which relies on analyzing the energy spectrum of 
recoils, will be affected by the precision to which \Leff~is known.

\section{Summary}
\label{sec:summary}

The work presented here represents a new measurement of \Leff~in an energy range
where it is poorly understood, but highly important to the field of dark matter
direct detection with liquid xenon detectors.  We show that at 10 keV and below,
this efficiency is lower than the average value of \Leff=$0.19$ while the
measurement in the literature \cite{Chepel:06} suggests a \emph{rise} in
\Leff~at these energies, albeit with large errors.  In light of the results of
our measurement, the XENON10 spin-independent limit is shifted up for WIMPs of
mass 100 GeV/c$^2$ by 12.5\%, while the high-mass regime is relatively
unchanged.

\begin{acknowledgments}
We wish to acknowledge the University of Florida High-Performance Computing 
Center for providing computational resources that have contributed to the 
simulations reported within this paper.  This work was funded by National 
Science Foundation grants No.~PHY-03-02646 and No.~PHY-04-00596, and by the 
Swiss National Foundation grant No.~20-118119.  Additional support was provided 
by grant No.~P41 EB002033-13, Radiological Research Accelerator Facility 
(RARAF), from the National Institutes of Health/National Institute of Biomedical 
Imaging and Bioengineering (NIBIB).  We express our gratitude to Dr.~Steve 
Marino of RARAF for the beam time and his support throughout the measurements.

\end{acknowledgments}

\end{document}